\documentclass[amsmath, amssymb, preprintnumbers,
aps,pra,superscriptaddress,onecolumn]{revtex4}

\usepackage{color} 
\usepackage{hyperref}
\usepackage{amsmath,amssymb,mathrsfs}
\usepackage{psfrag}
\usepackage{graphicx}
\usepackage{graphics}
\usepackage{subfigure}
\usepackage{epsfig}
\usepackage{bm}
\usepackage{verbatim,color,ulem}
\usepackage{braket}
\usepackage{verbatim}

\newcommand{\beq}{\begin{equation}}
\newcommand{\eeq}{\end{equation}}
\newcommand{\beqa}{\begin{eqnarray}}
\newcommand{\eeqa}{\end{eqnarray}}

\begin{document}

\title{Shift of the critical temperature in superconductors: \\ a self-consistent approach}
\author{Alberto Cappellaro}
\email{cappellaro@pd.infn.it}
\affiliation{Dipartimento di Fisica e Astronomia ``Galileo Galilei'',
	 Universit\`a di Padova, via Marzolo 8, 35131 Padova, Italy}
\author{Luca Salasnich}
\affiliation{Dipartimento di Fisica e Astronomia ``Galileo Galilei'',
Universit\`a di Padova, via Marzolo 8, 35131 Padova, Italy}
\affiliation{CNR-INO, via Nello Carrara, 1 - 50019 Sesto Fiorentino, Italy}

\begin{abstract}
Within the Ginzburg-Landau functional framework for the superconducting transition, 
we analyze the fluctuation-driven shift of the critical temperature. 
In addition to the order parameter fluctuations, we also take into account
the fluctuations of the vector potential above its vacuum.
We detail the approximation scheme to include the fluctuating fields contribution, 
based on the Hartree-Fock-Bogoliubov-Popov 
framework. We give explicit results for $d=2$ and $d=3$ spatial dimensions, in terms of
easily accessible experimental parameters such as the Ginzburg-Levanyuk 
number $\text{Gi}_{(d)}$, which is related to the width of the critical region where
fluctuations cannot be neglected, 
and the Ginzburg-Landau parameter $\kappa$, defined as the ratio between the magnetic
penetration length and the coherence one.
\end{abstract}

\maketitle

\section*{Introduction}

For over half a century, the Ginzburg-Landau (GL) theory has proven to be
the most effective theoretical tool to describe the critical behaviour 
of physical systems in presence
of symmetry breaking \cite{landau1937-1,ginzburg1950,kardar-fields}.
Despite its apparent simplicity, this functional approach 
can be successfully applied to a wide range of physical situations, 
such as, for instance neutral superfluids, spin systems and superconducting materials
\cite{kardar-fields,herbut-book,hohenberg-review}.
By relying on few well-grounded physical considerations (symmetries of the model and
the analyticity of thermodynamic functions), it provides a formidable platform
to understand the consequences of symmetry breaking and the role played by fluctuations
close to the criticality. 

Concerning conventional superconductors, it is well-known that fluctuations
do not play a relevant role and the mean-field analysis of the GL theory is 
extremely reliable in describing the superconducting transition. This is due to the fact
that the critical region cannot be accessed experimentally since it is very
narrow \cite{ginzburg1950}. As a consequence, certain questions,
such as the order of the transition 
or thermally-driven critical temperature shift,
remained a purely theoretical exercise \cite{halperin-1974} up to 
the first observation of high-T$_c$ superconductivity 
\cite{hightc}.  

Generally, many of these high-T$_c$ materials display a shorter coherence length $\xi$,
implying a higher Ginzburg-Levanyuk number $\text{Gi}_{(d)}$ and therefore a wider fluctuating
region, where a mean-field approach is not reliable \cite{ginzburg-1988,koo-2006}.
In particular, for a variety of materials,
fluctuations appear to be the main opponent of high-T$_c$ superconductivity. Since the
critical temperature shift is a non-universal property of physical systems, it is
important to properly understand how they affect, and eventually destroy, the 
superconducting phase \cite{sademelo-1993,perali-2002}. 
For instance, in multiband superconductors it has been found
that fluctuations can be suppressed by switching on a Josephson-like coupling  
between the bands \cite{salasnich-2019}.

In this paper we consider the minimal coupling of the usual $\psi^4$-theory with
the electromagnetic field. We aim to compute the fluctuation-drive shift of the critical
temperature (compared to the mean-field scheme) by taking into account
both the order parameter and the vector potential fluctuations. In order to 
perform this task we make use of an improved saddle-point equation, where fluctuations
are taken into account within the so-called Hartree-Fock-Bogoliubov-Popov scheme.
This approximation scheme is one of the many field-theoretical strategies to
model finite-temperature degenerate gases (for a review see \cite{shi-1998}),
where it has been proven to provide a reliable picture for bosonic alkali vapors
placed in an external confinement \cite{griffin-1996,minguzzi-1997,hutchinson-1997}. 
Here, this approach is applied to the Ginzburg-Landau equations, reading a modified
saddle-point equation where additional terms are present, depending on the
average of the square modulus of fluctuations. Once one has a strategy to self-consistently
compute these quantities, the critical temperature shift can be easily derived.

The paper is structured as follows: first, we review some general features of the
Ginzburg-Landau theory. In particular, we introduce the minimal coupling with the
electromagnetic field. Then we present our strategy to include the fluctuations in an
improved saddle-point (or Ginzburg-Landau) equation. We give explicit results for
$d=2$ and $d=3$ dimensions, where the shift turns out to depend only on 
the Ginzburg-Levanyuk number $\text{Gi}_{(d)}$ and the parameter 
$\kappa$, the latter being the ratio between the magnetic penetration length and the coherence one. 
Our analytical formulas for the shift of the critical temperature, directly obtained 
from the Ginzburg-Landau functional, are a generalization of familiar 
ones \cite{{larkin-book}}, which take into account only the fluctuations 
of the order parameter. 

\section*{Results}

\subsection*{The Ginzburg-Landau formulation: general features}

Concerning the normal-to-superconductor transition,
close to the critical temperature, the free
energy of a single-band superconductor can be split into
\begin{equation}
\mathcal{F} = \mathcal{F}_s + \mathcal{F}_n\;,
\end{equation}
with $\mathcal{F}_n$ the contribution of the disordered (i.e. normal) component, while
$\mathcal{F}_s$ models the contribution due to the emergence of an order phase 
(i.e. superconductor) 
characterized by an order parameter $\psi(\mathbf{r})$ acquiring a non-zero value
below a certain critical temperature to determine. 
The great intuition of Landau and Ginzburg consisted in writing down the latter in terms 
of few simple terms based on the symmetries to which the theory obeys. 
In our case, this implies 
\cite{larkin-book}
\begin{equation}
\begin{aligned}
\mathcal{F}_s[\psi,\mathbf{A}] = \int_{V}d^d\mathbf{r}\, \bigg[a(T)|\psi|^2 
+\frac{b}{2}|\psi|^4  
+ \gamma D_{\mu} \psi\,D_{\mu}^* \psi^* + \frac{(\nabla \wedge 
	\mathbf{A})^2}{2\mu_0}\bigg]
\end{aligned}
\label{gl-for-sc}
\end{equation}
where 
\begin{equation}
D_{\mu} = \partial_{\mu} - i e^* A_{\mu}(\mathbf{r})
\label{derivata gauge invariante}
\end{equation}
is the gauge-invariant derivative coupling the order parameter $\psi(\mathbf{r})$ with
the vector potential $\mathbf{A}(\mathbf{r})$, with $e^*$ being the effective charge.
In the equations above, the index $\mu$ labels the component of the vector $\mathbf{A}$.
One has also to recall that the order parameter
$\psi$ has to be assumed as a complex field if we aim to describe the 
normal-to-superconducting 
transition.

Concerning the couplings of the theory described by $\mathcal{F}_s$ 
in Eq. \eqref{gl-for-sc},
$b$ and $\gamma$  can be taken as positive constants, 
while the coefficient of the $|\psi|^2$-term contains the temperature
dependence. 
Simple considerations lead to the well-known conclusion that this dependence is linear, namely
\begin{equation}
a(T) = \alpha k_B\,(T-T_{c0})
\label{definizione del parametro a}
\end{equation}
with $\alpha>0$ and $k_B$ the Boltzmann constant. In the equation above,
$T_{c0}$ is the mean-field critical temperature.

It has to be remarked that Eq. \eqref{gl-for-sc} holds in absence of an 
external magnetic field.
In order to model this situation, $\mathcal{F}_s[\psi;\mathbf{A}]$ must be
modified adding a term proportional to $(\nabla \wedge A)\cdot \mathbf{H}$.
However, in this paper we limit our analysis to case of no
applied external field.

It is immediate to show that, 
with the ansatz $\psi(\mathbf{r}) = \psi_0$ and $\mathbf{A}(\mathbf{r}) = \bm{0}$,
one can find the usual saddle-point solution
\begin{equation}
|\psi^{\text{(mf)}}_0|^2 =
\begin{cases}
0 \quad \text{for $a(T) >0$}  \\
\\
-\dfrac{a(T)}{b}\quad \text{for $a(T) < 0$}\\
\end{cases}\;.
\label{saddle-point campo medio}
\end{equation}
Thus, the order parameter acquires a non-zero value when $a(T)$ changes its sign. 
Within the mean-field scheme,
the phase transition then occurs at a temperature given by the solution of 
\begin{equation}
a(T_{c0}) = 0\;.
\label{temperatura critica campo medio}
\end{equation}
This simple approach lacks every information about the fluctuation 
of the order parameter and the vector potential.
For conventional superconductors, the critical region where fluctuations 
are crucial is very small and cannot be observed. 

However, with the discovery of novel materials displaying exotic superconductivity, 
it appears clear that
deviations from the mean-field picture have to be explored very carefully.

\subsection*{Including the fluctuations: the HFBP scheme}
\subsubsection*{The saddle-point equation in the HFBP approximation}
Differently from the case of neutral superfluids, the Ginzburg-Landau functional 
for a superconducting system, as given by Eq. \eqref{gl-for-sc}, depends on two
different fields because of the minimal coupling with the electromagnetic field.

Thus, the saddle-point configuration of the system has to be determined by 
searching for the stationary trajectories of $\mathcal{F}_s[\psi,\mathbf{A}]$.
In other words, one has to solve the system
\begin{equation}
\dfrac{\delta \mathcal{F}_s}{\delta \psi^*} = 0 \quad\quad \text{and}
\quad\quad
\dfrac{\delta \mathcal{F}_s}{\delta {\mathbf{A}}} = 0
\end{equation}
where $\delta/\delta (\bullet)$ has to be intended in the sense of first variation.
The equation resulting from the variation with respect to $\psi^*$ reads
\begin{equation}
\bigg[a(T) + b|\psi|^2-\gamma\nabla^2 +\gamma (e^{*})^2\big|\mathbf{A}\big|^2
+2i\gamma e^*\mathbf{A}\cdot\nabla\bigg]\psi = 0\;,
\label{1st-var-psi}
\end{equation}
while $\delta \mathcal{F}_s / \delta{\mathbf{A}} = 0$ leads us to the following equation
\begin{equation}
\bigg[2\gamma(e^*)^2|\psi|^2 +
\frac{\nabla \wedge\nabla\wedge}{\mu_0}\bigg] \mathbf{A} =
ie^*\big( \psi\nabla\psi^* - \psi^*\nabla\psi\big)\;.
\label{1st-var-A}
\end{equation}
In both Eqs. \eqref{1st-var-psi} and \eqref{1st-var-A} the spatial dependence of both field 
is left intended. 
Obviously, assuming the ansatz $\psi(\mathbf{r}) = \psi_0$ and $\mathbf{A} = \mathbf{0}$ as in the
previous section, the Ginzburg-Landau equation are solved by 
Eq. \eqref{saddle-point campo medio}. It is then clear that a saddle-point configuration
exists for a uniform order parameter and in absence of a vector potential.

In order to encode the thermal fluctuations in a saddle-point scheme, let us then
split the order parameter
\begin{equation}
\psi(\mathbf{r}) = \psi_0 + \eta(\mathbf{r})
\label{split-psi}
\end{equation}
where $\psi_0$ is a constant which can be assumed as real but it is not necessarily 
given by $\psi_0^{\text{(mf)}}$ in Eq. \eqref{saddle-point campo medio}. On the other hand,
$\eta(\mathbf{r})$ is the space dependent fluctuation field.
The crucial feature concerning $\psi(\mathbf{r})$ as defined above is that
\begin{equation}
\psi_0 \equiv \braket{\psi(\mathbf{r})}
\label{broken-sym}
\end{equation}
and, as an immediate consequence,
\begin{equation}
\braket{\eta} = \braket{\eta^*} = 0\;.
\label{avg-fluct}
\end{equation}
The thermal average $\braket{\ldots}$ has to be intended as performed over a 
proper statistical ensemble where a global gauge symmetry can be spontaneously broken.

For the vector potential ${\bf A}({\bf r})$, we only consider
fluctuations ${\bf {\cal A}}({\bf r})$ above its vacuum ${\bf 0}$, namely
\begin{equation}
\mathbf{A}(\mathbf{r}) = \mathbf{0} + \bm{\mathcal{A}}(\mathbf{r})\;,
\label{vect-pot}
\end{equation}

Now, we proceed by replacing Eq. \eqref{split-psi} in Eq. \eqref{1st-var-psi}, such that
\begin{equation}
\begin{aligned}
a(T) \psi_0 + a(T) \eta +b\psi_0^3 +b\psi_0^2(2\eta + \eta^*) + b\psi_0(2|\eta|^2 + \eta^2)
& + b|\eta|^2\eta\\ 
-\gamma\nabla^2\eta + \gamma(e^*)^2|\bm{\mathcal{A}}|^2\psi_0 
& +\gamma(e^*)^2|\bm{\mathcal{A}}|^2\eta
+ 2i\gamma e^*\bm{\mathcal{A}}\cdot
\nabla\eta = 0\;.
\end{aligned}
\label{complicata-psi}
\end{equation}
We are interested in how fluctuations modify the uniform background 
$\psi_0$ within the broken symmetry phase, i.e. for $T < T_c$. We remark that, according 
to this framework, the critical temperature $T_c \neq T_{c0}$, since 
Eq. \eqref{temperatura critica campo medio} does not take into account the role 
of fluctuations.

In order to derive an equation for $\psi_0$ including a contribution due to fluctuations,
both of the order parameter and the vector potential, we take the thermal average
of Eq. \eqref{complicata-psi}. Because of Eq. \eqref{avg-fluct}, linear terms
in $\eta$ and $\eta^*$ are erased by default. One is then left with
\begin{equation}
\begin{aligned}
\bigg[a(T) + 2b\braket{|\eta|^2} + b\braket{\eta^2} +\gamma (e^*)^2
\braket{|\bm{\mathcal{A}}|^2}\bigg]\psi_0 +b\psi_0^3
+ b\braket{|\eta|^2 \eta} +\gamma (e^*)^2\braket{|\bm{\mathcal{A}}|^2\eta}
+2i\gamma e^* \braket{\bm{\mathcal{A}} \cdot \nabla \eta}
= 0\;,
\end{aligned}
\label{complicata-phi}
\end{equation}
which still appears rather complicated. Up to this point, we still have not fixed
the gauge for the vector potential. It is a well-known fact that the most natural choice
is the Coulomb (or transverse) gauge, where
\begin{equation}
\nabla \cdot \bm{\mathcal{A}} = 0\;.
\label{coulomb}
\end{equation}
Actually, besides the convenience matter, this fixing has profound physical consequences.
Indeed, it has been shown that only with Eq. \eqref{coulomb} the order parameter
correlations $\braket{\psi(\mathbf{r})\psi(0)}$ acquires a long-ranged character
(see, for instance, \cite{herbut-book}). 
Concerning  Eq. \eqref{complicata-phi}, let us notice that,
through a Fourier transformation, the Coulomb gauge is equivalently given by
$\tilde{\bm{\mathcal{A}}}(\mathbf{q})\perp \mathbf{q}$,
with $\mathbf{q}$ a wavevector in the reciprocal space. This implies that
$\braket{\bm{\mathcal{A}} \cdot \nabla \eta} = 0$.

Moving further, we simplify Eq. \eqref{complicata-phi} by means of the following 
approximation scheme. 
First, we neglect the three-field correlations, namely $\braket{|\eta|^2\eta}\simeq 0$ and
$\braket{|\bm{\mathcal{A}}|^2\eta}\simeq 0$. By drawing an analogy with the analysis of
bosonic gases at finite temperatures,
this corresponds to the Hartree-Fock-Bogoliubov (HFB) scheme \cite{griffin-1996}.
Moreover, we also discard the anomalous average, namely 
$\braket{\eta^2} = \braket{(\eta^*)^2} \simeq 0$, 
according to the 
so-called Popov approximation of the HFB framework (HFBP in the following). 

Thus, Eq. \eqref{complicata-phi} finally reads
\begin{equation}
\bigg[a(T) + 2b\braket{|\eta|^2} + \gamma (e^*)^2\braket{|\bm{\mathcal{A}}|^2}\bigg]\psi_0
+b\psi_0^3 = 0\;.
\label{gl-migliorata}
\end{equation}
whose solution, for $T < T_c$, is given by
\begin{equation}
\psi_0^2 = -\frac{a(T) +2b\braket{|\eta|^2} + \gamma 
	(e^*)^2\braket{|\bm{\mathcal{A}}|^2}}{b}\;.
\end{equation} 
From the equation is then immediate to derive the generalization of 
Eq. \eqref{temperatura critica campo medio} for the critical temperature, i.e.
\begin{equation}
a(T_c) + 2b\braket{|\eta|^2}_c + \gamma (e^*)^2\braket{|\bm{\mathcal{A}}|^2}_c = 0
\label{tcrit-migliorata}
\end{equation}
where $\braket{\ldots}_c$ implies that the average has to be computed at the critical 
point.

In contrast with Eq. \eqref{temperatura critica campo medio}, Eq. \eqref{tcrit-migliorata}
takes into account the presence of fluctuations both in the order parameter and
the vector potential, through the averages $\braket{|\eta|^2}$ and 
$\braket{|\bm{\mathcal{A}}|^2}$.

\subsubsection*{Averages in the HFBP scheme}
The shift of the critical temperature as 
in Eq. \eqref{tcrit-migliorata} requires the calculation of $\braket{|\eta|^2}$
and $\braket{|\bm{\mathcal{A}}|^2}$ at the criticality. 
As detailed in the Methods section, they can be computed by taking the GL equations
for the fluctuating fields $\eta(\mathbf{r})$ and $\bm{\mathcal{A}}(\mathbf{r})$ as
starting point (cfr. Eqs. \eqref{gl-eta} and \eqref{gl-delta}). 
From there, as outlined below, one can infer the corresponding Gaussian functional 
(i.e. a free energy, similarly to Eq. \eqref{gl-for-sc}) driving the thermodynamic properties
of fluctuations.
Indeed both Eq. \eqref{gl-eta} and Eq. \eqref{gl-delta} in Methods are the first variations
of a Gaussian functional. 
By working slightly above the critical temperature (i.e. $T \rightarrow T_c^{+}$), 
where $\psi_0 = 0$, Eq. \eqref{gl-eta} reads
\begin{equation}
\bigg[ a(T) +2b\braket{|\eta|^2} + \gamma (e^*)^2\braket{|\bm{\mathcal{A}}|^2}
-\gamma\nabla^2\bigg]\eta = 0\;,
\label{eta-above}
\end{equation}
which the trajectory stationarizing the functional
\begin{equation}
\begin{aligned}
\mathcal{F}^{(g)}_{\eta} [\eta,\eta^*]= \int_V d^d\mathbf{r}\; \eta^* \bigg[
a(T) &+2b\braket{|\eta|^2} + \gamma (e^*)^2\braket{|\bm{\mathcal{A}}|^2}
-\gamma\nabla^2\bigg]\eta
\end{aligned}
\label{funz-eta}
\end{equation}
with $V$ being the large $d$-dimensional volume enclosing the system.

Similarly, from Eq. \eqref{gl-delta}, one can infer the corresponding
functional for $\bm{\mathcal{A}}$, i.e.
\begin{equation}
\begin{aligned}
\mathcal{F}_{\bm{\mathcal{A}}}^{(g)}[\bm{\mathcal{A}}] = \int_V d^d\mathbf{r}\;
\bigg[
\gamma (e^*)^2\braket{|\eta|^2}|\bm{\mathcal{A}}|^2 +
\frac{(\nabla\wedge\bm{\mathcal{A}})^2}{2\mu_0}\;.
\bigg]
\end{aligned}
\label{funz-delta}
\end{equation} 
Each one of $\mathcal{F}^{(g)}_{\eta} [\eta]$ and 
$\mathcal{F}_{\bm{\mathcal{A}}}^{(g)}[\bm{\mathcal{A}}]$ are related to their corresponding
partition function, from which it is usually easy to compute average values and correlations,
since both of them are (functional) Gaussian integrals.

The presence of differential operators suggests that the calculation is easier in the 
Fourier space. Thus, with 
\begin{equation}
\eta(\mathbf{r}) = \frac{1}{\sqrt{V}} \sum_{\mathbf{q}}
e^{i\mathbf{q}\cdot\mathbf{r}}\;\tilde{\eta}(\mathbf{q})
\label{fourier-eta}
\end{equation}
and
\begin{equation}
\bm{\mathcal{A}}(\mathbf{r}) = \frac{1}{\sqrt{V}} \sum_{\mathbf{q}}
e^{i\mathbf{q}\cdot\mathbf{r}}\;\tilde{\bm{\mathcal{A}}}(\mathbf{q})
\label{fourier-delta}
\end{equation}
Eqs. \eqref{funz-eta} and \eqref{funz-delta} transform as, respectively,
\begin{equation}
\begin{aligned}
& \mathcal{F}^{(g)}_{\eta} =
\sum_{\mathbf{q}} \bigg[
a(T) + 2b\braket{|\eta|^2} + \gamma (e^*)^2\braket{|\bm{\mathcal{A}}|^2}+\gamma q^2
\bigg]|\tilde{\eta}(\mathbf{q})|^2 \\
\end{aligned}
\label{funz-eta-fourier}
\end{equation}
plus
\begin{equation}
\mathcal{F}^{(g)}_{\bm{\mathcal{A}}} = \sum_{\mathbf{q}} \bigg[
\gamma(e^*)^2\braket{|\eta|^2} + \frac{q^2}{2\mu_0}
\bigg]\big|\tilde{\bm{\mathcal{A}}}(\mathbf{q})\big|^2\;.
\label{funz-delta-fourier}
\end{equation}
Now, since both $\mathcal{F}^{(g)}_{\eta}[\tilde{\eta}]$ and 
$\mathcal{F}^{(g)}_{\bm{\mathcal{A}}}[\tilde{\bm{\mathcal{A}}}]$ are Gaussian, it is immediate
to infer that
\begin{equation}
\braket{|\eta|^2} = \frac{1}{\beta V} \sum_{\mathbf{q}}
\frac{1}{a(T) + 2b\braket{|\eta|^2} + \gamma (e^{*})^2\braket{|\bm{\mathcal{A}}|^2}
	+\gamma q^2}\;.
\label{eta2-avg}
\end{equation}
At the criticality Eq. \eqref{tcrit-migliorata} holds, therefore the equation
above is simplified into 
\begin{equation}
\braket{|\eta|^2}_c = \frac{k_B T_c}{V} \sum_{\mathbf{q}}\bigg(
\frac{1}{\gamma q^2}
\bigg)\;.
\label{eta2-avg-crit}
\end{equation}
Let us note that this result is the same one can derive in absence of
the minimal coupling with the electromagnetic field (i.e. $\mathbf{A} = \mathbf{0}$).
The crucial point is the fact that, on the contrary, $\braket{|\bm{\mathcal{A}}|^2}_c$
is affected by the fluctuation of the order parameter. Indeed, from 
Eq. \eqref{funz-delta-fourier}, one gets
\begin{equation}
\braket{|\bm{\mathcal{A}}|^2}_c = \frac{k_B T_c}{V} 
\sum_{\mathbf{q}} \frac{d-1}{2\gamma(e^*)^2\braket{|\eta|^2}_c + q^2/\mu_0}
\label{delta-avg-crit}
\end{equation}
where the factor $(d-1)$ is a consequence of the Coulomb gauge, setting to zero  
the component of $\tilde{\bm{\mathcal{A}}}(\mathbf{q})$ parallel to $\mathbf{q}$.
The vector potential then has only $(d-1)$ non-zero transverse components.

It worth to remember, at this point, the main results we have obtained by means
of the HFBP approximation scheme. First, we have derived the equation describing
the shift of the critical temperature (compared to the usual saddle-point result 
in Eq. \eqref{temperatura critica campo medio}). In order to actually solve 
Eq. \eqref{tcrit-migliorata}, 
we also need the average values (at the criticality) $\braket{|\eta|^2}_c$
and $\braket{|\bm{\mathcal{A}}|^2}_c$, respectively in 
Eqs. \eqref{eta2-avg-crit} and \eqref{delta-avg-crit}.

In the following, we are going to consider the continuum limit, 
namely (in spherical coordinates)
\begin{equation}
\sum_{\mathbf{q}} \rightarrow \frac{V}{(2\pi)^d}\,S_d\int_{q_0}^{\Lambda}
dq\; q^{d-1} \;,
\label{cont-lim}
\end{equation}
with $S_d= 2\pi^{d/2}/\Gamma(d/2)$ is the whole solid $d$-dimensional solid angle. 
In the equation above, we have introduced both an ultraviolet 
and an infrared cutoff to keep eventual divergences under control.

In the following section, we consider the case of $d=2$ and $d=3$ spatial dimensions.
An extremely interesting problem is represented by the dimensional crossover, 
i.e. the analysis of a thin film but with a finite thickness $\delta$. 
This physical realization has been investigated in presence of
a disordered environment
\cite{ovchinnikov-1973,takagi-1982,varlamov-1986}, reading an additional 
(logarithmic) shift of the 
critical temperature depending on $\delta$ and a properly defined diffusion coefficient.

\subsection*{Fluctuation-driven critical temperature shift}
\subsubsection*{The case $d=2$}
\begin{figure}[t!]
	\centering
	\includegraphics[width=0.7\columnwidth]{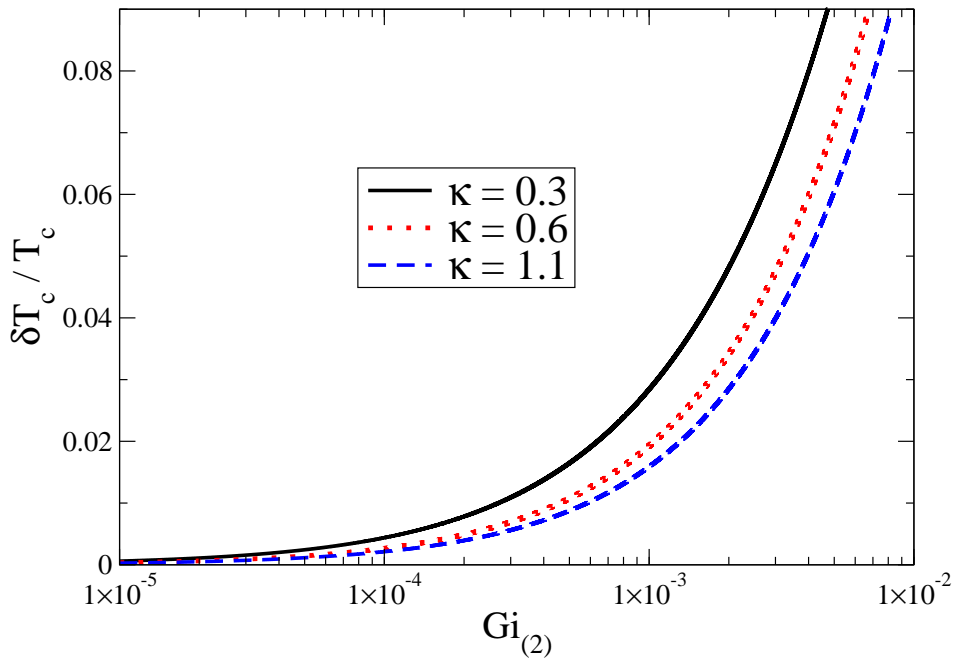}
	\caption{
		Shift of the critical temperature
		in two spatial dimensions
		according to Eq. \eqref{shift-2d} as a function
		of $\text{Gi}_{(2)}$ for three different values of $\kappa$. In the panel above,
		$\delta T_c = T_{c0} - T_c$.	
	}
	\label{fig2d}
\end{figure}

In $d=2$, by taking the continuum limit as in Eq. \eqref{cont-lim},
Eqs. \eqref{eta2-avg-crit} and \eqref{delta-avg-crit} easily lead us to
\begin{equation}
\braket{|\eta|^2}_c = \frac{k_B T_c}{2\pi\gamma} 
\;\ln\bigg(\frac{\Lambda}{q_0}\bigg)
\label{eta-d2}
\end{equation}
and 
\begin{equation}
\braket{|\bm{\mathcal{A}}|^2}_c = \frac{\mu_0 k_B T_c}{4\pi}
\; \ln\bigg(
\frac{\Lambda^2 +2\mu_0\gamma(e^*)^2\braket{|\eta|^2}_c}
{q_0^2 +2\mu_0\gamma(e^*)^2\braket{|\eta|^2}_c}
\bigg);.
\label{delta-d2}
\end{equation}
According to \cite{larkin-book}, a reasonable choice for the UV
cutoff is $\Lambda  \simeq 1/\xi_c$,
with 
\begin{equation}
\xi_c = \sqrt{\frac{\gamma}{\alpha k_B T_c}}\;.
\label{coherent}
\end{equation} 
Concerning the infrared cutoff $q_0$, we define it in terms of
the Ginzburg-Levanyuk number through
\begin{equation}
q_0 = \sqrt{\frac{\alpha k_B T_c}{\gamma}\, \text{Gi}_{(2)}}\;.
\label{infrared}
\end{equation}
The definition of the Ginzburg-Levanyuk number $\text{Gi}_{(d)}$ strongly depends
on the system dimensionality. Again, according to \cite{larkin-book}, in $d=2$ one has
\begin{equation}
\text{Gi}_{(2)} = \frac{b}{4\pi\gamma\alpha}\;.
\end{equation}
The UV cutoff is naturally defined by Eq. \eqref{coherent}, 
which specifies the minimal size of spatial fluctuations.		
The choice for the infrared cutoff in Eq. \eqref{infrared} is less
obvious and has to rely upon 
a renormalization-group (RG) argument \cite{salasnich-2019}. Within the
Wilson's standard framework of momentum-shell integration 
\cite{kardar-fields,herbut-book}, the flow equation for cutoff-dependent 
$a_{\lambda}(T)$ can be simplified by assuming the that the parameter $b$ does not
flow (i.e. it is equal to its bare value). 
The cutoff $\lambda$, separating the slow from fast modes in Wilson's approach, has an 
upper (i.e. ultraviolet) limit equal to $\lambda_{\infty} \rightarrow \Lambda = 1 /\xi_c$
with $\xi_c$ as in Eq. \eqref{coherent}.
In $d=2$, this approximation the equation for $a_{\lambda}(T)$
reads a logarithmic solution diverging when $\lambda \rightarrow 0$, 
naturally introducing the infrared cutoff $q_0$ as in Eq. \eqref{infrared}.
By solving simultaneously the RG equations for $a_{\lambda}$ and $b_{\lambda}$
the divergence disappears but, remarkably, it has been shown that the logarithmic
approximation outlined above works really well for $\text{Gi}_{(2)} \ll 1/2$,
which is exactly the range of values explored in this paper (cfr. Fig. \ref{fig2d}).

Therefore, by following the cutoff prescriptions outlined above,
Eqs. \eqref{eta-d2} and \eqref{delta-d2} result in
\begin{equation}
\braket{|\eta|^2}_c = \frac{k_BT_c}{4\pi\gamma}\ln\bigg(
\frac{1}{\text{Gi}_{(2)}}
\bigg)
\label{eta-finale-2d}
\end{equation}
and
\begin{equation}
\braket{|\bm{\mathcal{A}}|^2}_c =
\frac{\mu_0 k_B T_c}{4\pi} \ln\bigg[
\frac{1+ \frac{\text{Gi}_{(2)}}{\kappa^2}\ln\big(\frac{1}{\text{Gi}_{(2)}}\big)}
{\text{Gi}_{(2)}+ \frac{\text{Gi}_{(2)}}{\kappa^2}\ln\big(\frac{1}{\text{Gi}_{(2)}}\big)}
\bigg]\;.
\label{delta-finale-2d}
\end{equation}
The parameter $\kappa$ is the usual ratio between the magnetic penetration length $\lambda(T)$
and the coherence length $\xi(T)$, i.e. $\kappa = \lambda(T)/\xi(T)$. We recall that $\kappa$ 
is a crucial quantity in the GL theory for superconductors, since type-I superconductors
all have $\kappa < 1/\sqrt{2}$. On the contrary, the condition $\kappa > 1/\sqrt{2}$ characterizes
type-II superconducting materials.
By assuming
\begin{equation}
\xi(T) =\sqrt{\frac{\gamma}{a(T)}}\quad \text{and} \quad 
\lambda(T)=\sqrt{\frac{b}{2\mu_0(e^*)^2\gamma a(T)}} \;,
\label{lunghezze}
\end{equation}  
one easily gets
\begin{equation}
\kappa = \frac{\lambda(T)}{\xi(T)} = 
\sqrt{\frac{b}{2\mu_0(e^*)^2\gamma^2} }  \;.
\label{kappa}
\end{equation}
It has to be remarked that the equation above still holds for $d=3$.

By replacing Eqs. \eqref{eta-finale-2d} and \eqref{delta-finale-2d} in
Eq. \eqref{tcrit-migliorata}, the shift of the critical temperature in $d=2$, compared to
the usual saddle-point picture, is given by
\begin{equation}
\begin{aligned}
\frac{T_{c0}-T_c}{T_c} &= 2\text{Gi}_{(2)}\ln\bigg(\frac{1}{\text{Gi}_{(2)}}\bigg)+ \frac{\text{Gi}_{(2)}}{2\kappa^2} \ln\bigg[
\frac{1+ \frac{\text{Gi}_{(2)}}{\kappa^2}\ln\big(\frac{1}{\text{Gi}_{(2)}}\big)}
{\text{Gi}_{(2)}+ \frac{\text{Gi}_{(2)}}{\kappa^2}\ln\big(\frac{1}{\text{Gi}_{(2)}}\big)}
\bigg]\;.
\end{aligned}
\label{shift-2d}
\end{equation}
We stress that only two parameters, 
the Ginzburg-Levanyuk number $\text{Gi}_{(2)}$ and the Ginzburg-Landau ratio $\kappa$,
drive the shift of the critical temperature in two spatial dimensions. 
Eq. (\ref{shift-2d}) is one on the main results of the paper. 
The first addend of this equation, already known in literature \cite{salasnich-2019,larkin-book}, 
takes into account the thermal fluctuations of the order parameter . 
The second addend takes instead into account the thermal 
fluctuations of the electromagnetic vector potential around its vacuum. 
In fig. \ref{fig2d} we report the behaviour of the critical temperature shift given
by Eq. \eqref{shift-2d}
(with $\delta T_c = T_{c0}-T_c$) as a function
of $\text{Gi}_{(2)}$ for three different values of $\kappa$. 
Let us remark that the displacement from the mean-field prediction $T_{c0}$
is progressively reduced by moving towards to the type-II regime.
%

It is important to recall that two-dimensional physical systems require from us to also
consider the eventual occurring of the Berezinskii-Kosterlitz-Thouless (BKT) transition
\cite{berezkinskii-1972,kosterlitz-1973}. Its peculiar behaviour
can be understood in terms of phase fluctuations of the order parameter.
Indeed, a complex field can be naturally characterized in terms of two real fields by
means of the phase-amplitude representation
\begin{equation}
\psi(\mathbf{r}) = \Psi_{0}\; \exp\big\lbrace i\phi(\mathbf{r})\big\rbrace\;.
\label{amplitude-phase representation}
\end{equation}
In the equation above, $\Psi_{0}$ can be taken as the uniform solution of the	
GL equations. By neglecting the crucial role played by the phase field $\phi(\mathbf{r})$,
the amplitude acquires a non-zero value for $T < T_c$, with $T_c$ being the 
shifted (compared to the mean-field result) critical temperature in Eq. \eqref{shift-2d}.
On the other hand, the phase field $\phi(\mathbf{r})$ is obviously defined on a compact
support. As a consequence, 
the system can display nontrivial topological excitations
in form of quantized vortices.

In Methods we outline a procedure providing, in first approximation,
the additional BKT shift \cite{larkin-book,salasnich-2019}
by means of the Nelson-Kosterlitz criterion \cite{nelson-1977}.
According to this approach, $T_{BKT}$ is shifted with respect to the unrenormalized
contribution in Eq. \eqref{shift-2d} by
\begin{equation}
\frac{T_c - T_{BKT}}{T_{BKT}} = 4\; \text{Gi}_{(2)}\;.
\label{bkt-shift}
\end{equation}
It is important to remark that, 
while the Eq. \eqref{bkt-shift} has a very simple appearance,
the fundamental input is the critical temperature computed in Eq. \eqref{shift-2d}.

\subsubsection*{The case $d=3$}
\begin{figure}[ht!]
	\centering
	\includegraphics[width=0.7\columnwidth]{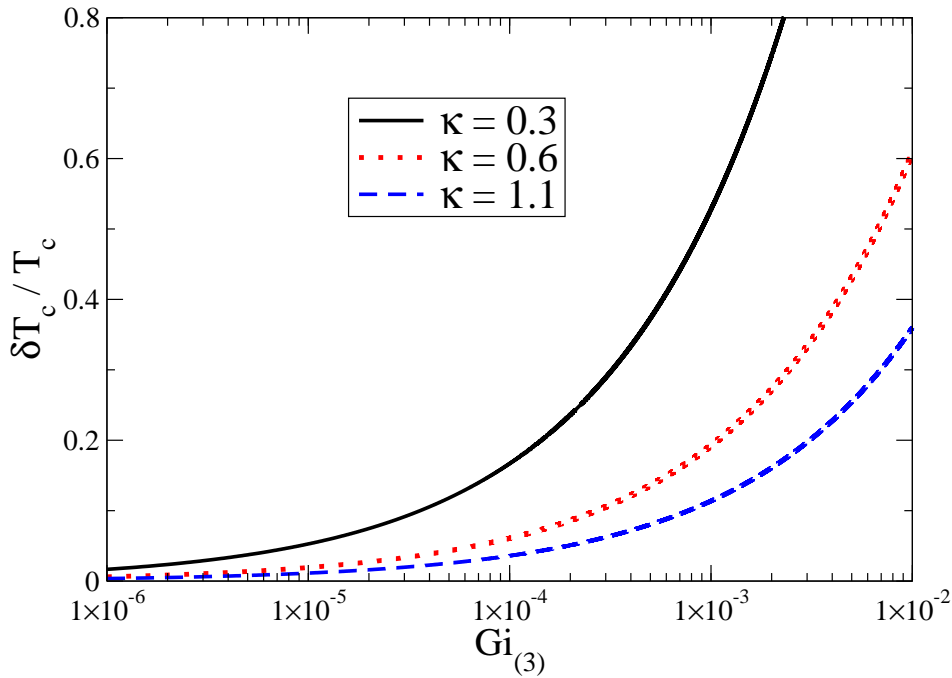}
	\caption{
		Shift of the critical temperature
		in three spatial dimensions
		according to Eq. \eqref{shift-3d} as a function
		of $\text{Gi}_{(3)}$ for different values of $\kappa$. 
		For the sake of comparison we have made use of the same values of fig. \ref{fig2d}.
		In the panel above,
		$\delta T_c = T_{c0} - T_c$.		
	}
	\label{fig3d}
\end{figure}

In the case $d=3$, the integration over $q$ as prescribed by Eq. $\eqref{cont-lim}$
of Eqs. \eqref{eta2-avg-crit} and \eqref{delta-avg-crit}
gives us back
\begin{equation}
\braket{|\eta|^2}_c = \frac{k_B T_c \Lambda}{2\pi^2\gamma}
\label{eta-d3}
\end{equation}
and
\begin{equation}
\begin{aligned}
\braket{|\bm{\mathcal{A}}|^2}_c  = \frac{\mu_0k_B T_c}{\pi^2}\bigg[
\Lambda - \sqrt{2\mu_0(e^*)^2\gamma \braket{|\eta|^2}_c}  \arctan\bigg(
\frac{\Lambda}{\sqrt{2\mu_0(e^*)^2\gamma \braket{|\eta|^2}_c}}
\bigg)
\bigg]\;.
\end{aligned}
\label{delta-d3}
\end{equation}
No infrared divergence arises by performing the integration leading to the equations above,
so we have safely taken the limit $q_0 \rightarrow 0$.
We also approximate the $\arctan(\ldots) \rightarrow \pi/2$ since it provides
only subleading contribution, compared to the coefficient in front of it.
However, an ultraviolet divergence are still 
present: as for $d=2$, we assume $\Lambda \simeq 1/\xi_c$, with $\xi_c$
given by Eq. \eqref{coherent}, holding also in three spatial dimensions.
The Ginzburg-Levanyuk number, on the contrary, reads \cite{larkin-book}
\begin{equation}
\text{Gi}_{(3)} = \frac{b^2}{64\pi^2 \alpha \gamma^3} k_B T_c\;.
\end{equation}
Consequently,
\begin{equation}
\braket{|\eta|^2}_c = \frac{4\alpha k_B T_c}{\pi b} \sqrt{\text{Gi}_{(3)}}
\label{eta-final-3d}
\end{equation}
and
\begin{equation}
\begin{aligned}
\braket{|\bm{\mathcal{A}}|^2}_c = \frac{\mu_0 k_B T_c}{\pi^2}\bigg[
\sqrt{\alpha k_B T_c}{\gamma} 
- \sqrt{2\pi\mu_0(e^*)^2\alpha\gamma \sqrt{\text{Gi}_{(3)}}\;\frac{k_B T_c}{b}}\bigg]\;.
\end{aligned}
\label{delta-final-3d}
\end{equation}
By inserting Eqs. \eqref{eta-final-3d} and \eqref{delta-final-3d} in
Eq. \eqref{tcrit-migliorata}, up to the leading term in $\text{Gi}_{(3)}$,
the shift of the critical temperature results in 
\begin{equation}
\begin{aligned}
\frac{T_{c0}-T_c}{T_c}  = \frac{8}{\pi} \sqrt{\text{Gi}_{(3)}} + 
\frac{4}{\pi \kappa^2} \sqrt{\text{Gi}_{(3)}}
\end{aligned}
\label{shift-3d}
\end{equation}
with $\kappa$ defined in Eq. \eqref{kappa}. 
Similarly to the $d=2$ case, in fig. \ref{fig3d} we plot
Eq. \eqref{shift-3d} as a function of the Ginzburg-Levanyuk number for 
different values of the ratio $\kappa$. Again, we see that, at higher values
of $\kappa$, namely for higher penetration lengths of lower coherence ones, 
$(T_{c0}-T_c)/T_c$ is remarkably reduced.

Again, by assuming the minimal coupling with the electromagnetic field, 
the Ginzburg-Levanyuk number 
$\text{Gi}_{(3)}$ and the Ginzburg-Landau ratio $\kappa$ are all one needs 
to characterize the critical temperature shift from the usual mean-field
scheme. Eq. (\ref{shift-3d}) is the another important result of our paper. 
The first addend of this equation is already known in 
literature \cite{kardar-fields,larkin-book} while the second addend, related to 
the fluctuations of the vector potential, is instead a novel result. 

\section*{Discussion}

In this paper, we have presented a self-consistent approach to include the fluctuations
of order parameter and vector potential in an improved saddle-point equation.
As a consequence, it has been possible to derive the fluctuation-driven shift of
$T_c$ both in $d=3$ and $d=2$. Remarkably, this displacement from the usual mean-field
results only depends from two combinations of the GL coupling constants: 
the Ginzburg-Levanyuk number $\text{Gi}_{(d)}$, related to the width of the 
critical region, and the ratio $\kappa = \lambda(T)/\xi(T)$, dividing type-I from type-II
superconductors.

In presence of a wide critical region,
namely a high value of $\text{Gi}_{(d)}$, it is crucial to properly consider
not only the role of fluctuations on the various thermodynamic functions, but also
the eventual strategies apt to control them. This is not a purely academical question;
on the contrary it can have relevant technological applications, since
this kind of materials are the ones displaying pairing mechanisms leading
to high-temperature superconductivity, such as electron-hole superfluidity
\cite{hirsch-1990,hirsch-1991,perali-2018}.  
Indeed, as already mentioned, a inter-band Josephson-like coupling has been proved to
contain the detrimental effect of fluctuations and 
preserve an optimal superconductivity regime \cite{salasnich-2019}, 
also in reduced dimensionality, where coherent behaviour is otherwise greatly suppressed
\cite{perali-2020}. 

By adopting this point of view,
in this paper we have considered a single-band structure but in presence of
a minimal coupling to the electromagnetic field. As a consequence, it turns
out the fluctuations of the vector potential $\bm{\mathcal{A}}$ that the critical
temperature acquires an additional shift depending from $\kappa$ 
(cfr. Eqs. \eqref{shift-2d} and \eqref{shift-3d}). In other words, it depends
on how much the material sample can be penetrated by the magnetic fields. 
A natural extension of this analysis is involving the presence of at least another
band with a proper coupling, together the possibility to introduce the particle statistics
moving to a quantum framework which is closer, for instance, to the phenomenology of the
BCS-BEC crossover \cite{sademelo-1993,perali-2019}.

\section*{Methods}

\subsection*{The GL equations for the fluctuating fields}
The computation of the shifted critical temperature
then requires the self-consistent calculation of these averages.

In this section we detail the derivation of the GL equations for the fluctuations
of the order parameter and the vector potential, namely $\eta(\mathbf{r})$ and
$\bm{\mathcal{A}}(\mathbf{r})$. These equations then serve as starting point to
engineer a (Gaussian) functional accounting for the statistical properties of the 
fluctuating fields.
This task can be fulfilled by looking back at Eq. \eqref{complicata-psi}. 
Here, the HFB scheme prescribes that all the terms more than linear in the fluctuating
fields are decoupled with a couple-by-couple average. For instance, 
$|\eta|^2\eta \simeq 2\braket{|\eta|^2}\eta + \braket{\eta^2}\eta^*$, 
$|\eta|^2 \simeq \braket{|\eta|^2}$ and 
$|\bm{\mathcal{A}}|^2\eta \simeq \braket{|\bm{\mathcal{A}}|}\eta + 
2\braket{\eta\bm{\mathcal{A}}}\cdot\bm{\mathcal{A}}$. In addition, the  Popov prescription
imposes $\braket{\eta^2} \simeq 0$. We also assume that the order parameter and the 
vector potential fluctuates independently, so
\begin{equation}
\braket{\eta\;\delta A_{\mu}} = \braket{\eta}\braket{\delta A_{\mu}} = 0
\label{stat-ind}
\end{equation}
where the last equality is due to Eq. \eqref{avg-fluct}.
This series of considerations leads us to the GL equation for the order parameter
fluctuations,
\begin{equation}
\begin{aligned}
\bigg[
a(T) +2b\braket{|\eta|^2} & + \gamma (e^*)^2\braket{|\bm{\mathcal{A}}|^2} \eta 
+ 2 b \psi_0^2  -\gamma\nabla^2
\bigg] + b\psi_0^2\eta^*   =0 \;.
\end{aligned}
\label{gl-eta}
\end{equation}
The corresponding equation for $\bm{\mathcal{A}}$ can be easily derived
by following the same steps detailed above, together with the Coulomb gauge. 
More in detail, by replacing Eqs. \eqref{split-psi} and \eqref{vect-pot}
in \eqref{1st-var-A}, together with the assumption in Eq. \eqref{stat-ind},
this procedure transforms the left-hand-side (LHS) as
\begin{equation}
\begin{aligned}
\bigg[2\gamma(e^*)^2|\psi|^2 +
\frac{\nabla \wedge\nabla\wedge}{\mu_0}\bigg] \mathbf{A}
& \simeq 
2\gamma (e^*)^2(\psi_0^2+\braket{|\eta|^2}) \bm{\mathcal{A}}
+ \frac{\nabla\wedge\nabla\wedge \bm{\mathcal{A}}}{\mu_0}\;.
\end{aligned}
\label{lhs}
\end{equation}
Concerning the right-hand-side of Eq. \eqref{1st-var-A}, additional comments are in order.
First, let us note that, because of Eq. \eqref{split-psi}, terms linear in $\nabla \eta$
and $\nabla\eta$ arise. They are linear in $\mathbf{A}$ in the Ginzburg-Landau functional
whose Eq. \eqref{1st-var-A} is a first variation. Because of the Coulomb gauge, terms 
of the kind $\bm{\mathcal{A}} \cdot \nabla\eta$ give a null contribution to the functional
and consequently should not appear in the corresponding GL equation.
The remaining term is simply
\begin{equation}
\eta^*\nabla\eta - \eta\nabla\eta^* \;.
\end{equation}
Since $\eta$ is complex, it can be equivalently rephrased as
\begin{equation}
\eta^*\nabla\eta - \eta\nabla\eta^* = 2i\big(\text{Im}\,\eta \nabla\text{Re}\,\eta -
\text{Im}\,\eta\nabla\text{Re}\,\eta\big)\;.
\end{equation}
The last approximation we perform, together with the HFBP scheme and
Eq. \eqref{stat-ind}, is the statistical independence of the real and imaginary part of
$\eta$. It can be shown that, up to the Gaussian order in the fluctuating fields,
within the usual perturbative scheme \cite{kardar-fields}, this assumption holds.
In the end, no contribution comes from the right-hand-side of Eq. \eqref{1st-var-A}, therefore one
is left with the following equation for $\bm{\mathcal{A}}$
\begin{equation}
\bigg[2\gamma (e^*)^2 \big( \psi_0^2 +\braket{|\eta|^2}\big) + 
\frac{1}{\mu_0}\nabla\wedge\nabla\wedge\bigg]\bm{\mathcal{A}} = 0 \;.
\label{gl-delta}
\end{equation}
In the end, within the HFBP we have derived three GL equations, one for
the homogeneous background $\psi_0$, Eq. \eqref{gl-migliorata}, the other two for
$\eta$ and $\bm{\mathcal{A}}$, respectively Eqs. \eqref{gl-eta} and \eqref{gl-delta}.

\subsection*{Including the BKT contribution}
As a starting point, we take the field $\psi(\mathbf{r})$ in its phase-amplitude
representation given by Eq. \eqref{amplitude-phase representation}.
In order to compute the additional BKT shift, 
as a first approximation\cite{larkin-book,salasnich-2019} one can consider the
following phase-only functional
\begin{equation}
\mathcal{F}[\phi] = \mathcal{F}_0 + 
\frac{J(T)}{2}\int d^2\mathbf{r} |\nabla\phi(\mathbf{r})|^2 \;.
\label{phase-only functional}
\end{equation}
The phase stiffness $J(T)$ is defined in terms of the original GL parameters, i.e. 
\begin{equation}
J(T) = \frac{2\gamma\alpha}{b} \,k_B (T-T_c) \;.
\label{sf stiffness}
\end{equation}
The key point in the Eqs. \eqref{phase-only functional} and \eqref{sf stiffness}
consists in the \textit{benchmark} temperature in 
$J(T)$: it is no more $T_{c0}$ as for $a(T)$ 
in Eq. \eqref{definizione del parametro a}, but the 
fluctuation-shifted $T_c$ in Eq. \eqref{shift-2d}. 

As already clear from the seminal papers 
\cite{berezkinskii-1972,kosterlitz-1973}, the major step forward 
in the BKT understanding was the fact that it is actually a topological transition. 
Indeed, the compactness of the phase field $\phi(\mathbf{r})$ implies the
possibility for the system to display excited configurations (compared to the uniform
one) which cannot be reached by continuously deforming the order parameter
\cite{kardar-fields}. In $d=2$ these excitations are simply vortices
and antivortices, depending on the sign of their (topological) charge 
$\nu \in \mathrm{Z}$, 
defined in terms of phase winding
\begin{equation}
\frac{1}{2\pi}\oint_{\Gamma} d^2\mathbf{r}\; \nabla\phi(\mathbf{r}) = \nu \;.
\label{vortex charge}
\end{equation}
Now, a standard approach consists in drawing an electrostatic analogy and treating
vortices and antivortices as a Coulomb gas moving in a uniform and neutral background
\cite{kardar-fields,herbut-book,minnhagen-1987}. The BKT critical point then 
divides two different phase of this vortex gas: for $T< T_{BKT}$, bound states
of vortex-antivortex are predominant, while they are unbound 
above the critical temperature,
destroying every global coherence property, such as superfluidity.
Remarkably, this appears in a discontinuous way. Indeed, the transition displays
a universal jump in the phase stiffness (and, consequently, in the superfluid
density). This peculiar feature of the transition can be used to compute the
critical temperature, according to the so-called Nelson-Kosterlitz criterion
\cite{kosterlitz-1973}
\begin{equation}
T_{BKT} = \frac{\pi}{2}\,J(T_{BKT})\;
\label{nelson criterion}
\end{equation}
reading Eq. \eqref{bkt-shift} in the main text.

\bibliography{References}

\begin{thebibliography}{999}
	
	\bibitem{landau1937-1} Landau, L. D.
	\textit{On the theory of phase transition},
	Zh.Eksp.Teor.Fiz. \textbf{7}, 19 (1937); Landau, L. D. 
	\textit{On the theory of phase transitions. II.},
	\textit{ibid.}, 627 (1937).
	
	\bibitem{ginzburg1950}	Ginzburg, V. L. and Landau, L. D. 
	\textit{On the Theory of superconductivity},
	Zh. Elsp. Teor. Fiz. \textbf{20}, 1064
	(1950).
	
	\bibitem{kardar-fields} Kardar, M. 
	\textit{Statistical Physics of Fields}
	(Cambridge University Press, Cambridge, 2007).
	
	\bibitem{herbut-book} Herbut, I. \textit{A Modern Approach to Critical Phenomena}
	(Cambridge University Press, New York, 2010).
	
	\bibitem{hohenberg-review} Hohenberg, P. C. and Krekhov, A. P.
	\textit{An introduction to the Ginzburg-Landau theory of phase transition and
		non-equilibrium patterns}, Phys. Rep. \textbf{572}, 1-42 (2015).
	
	\bibitem{halperin-1974} Halperin, B. I. Lubensky, T. C. and Ma, S.-K.
	\textit{First-Order Phase Transitions in Superconductors and Smectic-A Liquid Crystals},
	Phys. Rev. Lett. \textbf{32},292 (1974).
	
	\bibitem{hightc} Chu, C. W. \textit{et al.},
	\textit{Evidence for superconductivity above 40 K in the La-Ba-Cu-O compound system},
	Phys. Rev. Lett. \textbf{58}, 405 (1987);
	Cava, R. J, van Dover, R. B., Batlogg, B. and Rietman, E. A.,
	\textit{Bulk superconductivity at 36 K in La$_{1.8}$Sr$_{0.2}$CuO$_4$},
	Phys. Rev. Lett. \textbf{58}, 408 (1987);
	Bednorz, J. G. and Muller, K. A.
	\textit{Perovskite-type oxides—The new approach to high-T$_c$ superconductivity},
	Rev. Mod. Phys.\textbf{60}, 585 (1988).
	
	\bibitem{ginzburg-1988} Bulaevskii, L. N., Ginzburg, V. L. and Sobyanin, A. A.
	\textit{Macroscopic theory of superconductors with small coherence length},
	Zh. Eksp.  Teor. Fiz. \textbf{94}, 355-375 (1988).
	
	\bibitem{koo-2006} Koo, J.-H. and Cho, G.
	\textit{Short coherence length in cuprate superconductors},
	Mod. Phys. Lett. B \textbf{20},539-547 (2006).
	
	\bibitem{sademelo-1993} S\'a de Melo, C., Randeria, M. and Engelbrecht, J.
	\textit{Crossover from BCS to Bose superconductivity: transition temperature and
		time-dependent Ginzburg-Landau theory},
	Phys. Rev. Lett. \textbf{71}, 3202 (1993).
	
	\bibitem{perali-2002} Perali, A., Strinati, G. C., and Castellani, C.
	\textit{Pseudogap and spectral function from superconducting fluctuations 
		to the bosonic limit},
	Phys. Rev. B \textbf{66}, 024510 (2002).
	
	\bibitem{salasnich-2019} Salasnich, L., Shanenko, A. A., Vagov, A., Albino Aguiar, J.
	and Perali, A.
	\textit{Screening of pair fluctuations in superconductors with coupled shallow
		and deep bands: a route to higher temperature superconductivity},
	Phys. Rev. B \textbf{100}, 64510 (2019).
	
	\bibitem{shi-1998} Shi, H. and Griffin, A.
	\textit{Finite-temperature excitations in a dilute Bose-condensed gas},
	Phys. Rep. \textbf{304}, 1-87 (1998).
	
	\bibitem{griffin-1996} Griffin, A. \textit{Conserving and gapless approximations 
		for an inhomogeneous Bose gas at finite temperatures}, Phys. Rev. B \textbf{53}, 9341 
	(1996).
	
	\bibitem{minguzzi-1997} Minguzzi, A. and Tosi, M. P.
	\textit{Linear density response in the random phase approximation 
		for confined Bose vapours at finite temperature},
	J. Phys.: Condens. Matter \textbf{9}, 12011 (1997).
	
	\bibitem{hutchinson-1997} Hutchinson, D. A. W., Zaremba, E. and Griffin, A.
	\textit{Finite Temperature Excitations of a Trapped Bose Gas},
	Phys. Rev. Lett. \textbf{78}, 1842 (1997).
	
	\bibitem{larkin-book} Larkin, A. and Varlamov, A. 
	\textit{Theory of Fluctuations in Superconductors} 
	(Oxford University Press, Oxford, 2005).
	
	\bibitem{ovchinnikov-1973} Ovchinnikov, Y. N.
	\textit{Fluctuation shift of the transition temperature of thin superconductors},
	Zh. Eksp. Teor. Fiz. \textbf{64}, 719-724 (1973).
	
	\bibitem{takagi-1982} Takagi, H. and Kuroda, Y.
	\textit{Anderson localization and superconducting transition temperature in two
		dimensional systems}, Solid State Communications \textbf{41}, 643-648 (1982).
	
	\bibitem{varlamov-1986} Varlamov, A. A. and Dorin, V. V.
	\textit{Influence of electronelectron interaction on the critical current in a Josephson
		junction},
	Zh. Eksp. Teor. Fiz. \textbf{91}, 1955 (1986).
	
	\bibitem{berezkinskii-1972}  V.L. Berezinskii, 
	\textit{Destruction of Long-range Order in One-dimensional and
		Two-dimensional Systems having a Continuous Symmetry Group I. Classical Systems},
	Sov. Phys. JETP \textbf{32}(3), 493-500 (1972); V. L. Berezkinskii,
	\textit{Destruction of long-range order in one-dimensional and two-dimensional 
		systems with a continuous symmetry group. II. Quantum systems}, 
	\textit{ibid.} \textbf{34}(3), 610-616 (1972).
	
	\bibitem{kosterlitz-1973} J. M. Kosterlitz and D. J. Thouless,
	\textit{Ordering, metastability and phase transitions in two-dimensional systems},
	J. Phys. C: Solid State Phys. \textbf{6}, 1181 (1973).
	
	\bibitem{nelson-1977} D. R. Nelson and J. M. Kosterlitz,
	\textit{Universal Jump in the Superfluid Density of Two-Dimensional Superfluids},
	Phys. Rev. Lett. \textbf{39}, 1201 (1977).
	
	
	\bibitem{hirsch-1990} Marsiglio, F. and Hirsch, J. E.
	\textit{Hole superconductivity and the high-T$_c$ oxides},
	Phys. Rev. B \textbf{41}, 6435 (1990).
	
	\bibitem{hirsch-1991}  Hirsch, J. E. and Marsiglio, F.
	\textit{Hole superconductivity in oxides: A two-band model},
	Phys. Rev. B \textbf{43}, 424 (1991).
	
	\bibitem{perali-2018} Saberi-Pouya, S., Zarenia, M., Perali, A., Vazifehshenas, T. and
	Peeters, F. M.
	\textit{High temperature electron-hole superfluidity with 
		strong anisotropic gaps in double phosphorene monolayers},
	Phys. Rev. B \textbf{97}, 174503 (2018).
	
	\bibitem{perali-2020} Saraiva, T. T., Cavalcanti, P. J. F., Vagov, A.
	Vasenko, A. S., Perali, A., Dell'Anna, L. and Shanenko, A. A.
	\textit{Quasi-one-dimensional system as a high-temperature superconductor},
	arXiv:2002.01989 (2020).
	
	\bibitem{perali-2019} Tajima, H., Yerin, Y., Perali, A. and Pieri, P.
	\textit{Enhanced critical temperature, pairing fluctuation effects, 
		and BCS-BEC crossover in a two-band Fermi gas}, 
	Phys. Rev. B \textbf{99}, 180503 (2019).
	
	\bibitem{minnhagen-1987} P. Minnhagen,
	\textit{The two-dimensional Coulomb gas, vortex unbinding,
		and superfluid-superconducting films},
	Rev. Mod. Phys. \textbf{59}, 1001 (1987).
	
\end{thebibliography}

\textbf{Acknowledgements} The authors thank A. Varlamov and A. Perali for their
constructive comments on this manuscript,
and M. Taffarello for the insightful discussions during the initial stage of this work.
\\

\textbf{Author contributions statement} Both authors equally contributed to this work.
\\

\textbf{Additional information}
The authors declare no conflict of interests.

\end{document}